\RequirePackage{fix-cm}
\documentclass[]{pasj01}

\Received{2022/08/19}
\Accepted{2022/11/19}

\usepackage{anyfontsize}
\usepackage{mathpazo}
\usepackage[T1]{fontenc}
\usepackage{makecell, multirow}
\usepackage{tabularx, booktabs}
\usepackage{dblfloatfix}
\usepackage{url}

\begin{document}

\title{
Detection of a bright burst from the repeating FRB 20201124A at 2 GHz }

\author{Sota \textsc{Ikebe}\altaffilmark{1,2}}
\email{ikebe-souta615@g.ecc.u-tokyo.ac.jp}
\author{Kazuhiro \textsc{Takefuji}\altaffilmark{3}}
\author{Toshio \textsc{Terasawa}\altaffilmark{4}}
\author{Sujin \textsc{Eie}\altaffilmark{1,2,5}}
\author{Takuya \textsc{Akahori}\altaffilmark{2,6}}
\author{Yasuhiro \textsc{Murata}\altaffilmark{3,7,8}}
\author{Tetsuya \textsc{Hashimoto}\altaffilmark{9}}
\author{Shota \textsc{Kisaka}\altaffilmark{10}}
\author{Mareki \textsc{Honma}\altaffilmark{1,2,11}}
\author{Shintaro \textsc{Yoshiura}\altaffilmark{2}}
\author{Syunsaku \textsc{Suzuki}\altaffilmark{2}}
\author{Tomoaki \textsc{Oyama}\altaffilmark{11}}
\author{Mamoru \textsc{Sekido}\altaffilmark{12}}
\author{Kotaro \textsc{Niinuma}\altaffilmark{13,14}}
\author{Hiroshi \textsc{Takeuchi}\altaffilmark{7,8}}
\author{Yoshinori \textsc{Yonekura}\altaffilmark{15}}
\author{Teruaki \textsc{Enoto}\altaffilmark{16,17}}

\altaffiltext{1}{Department of Astronomy, School of Science, The University of Tokyo, 7-3-1 Hongo, Bunkyo-ku, Tokyo 113-0033, Japan}
\altaffiltext{2}{Mizusawa VLBI Observatory, National Astronomical Observatory of Japan, 2-21-1 Osawa, Mitaka, Tokyo 181-0015, Japan}
\altaffiltext{3}{Usuda Deep Space Center, Japan Aerospace Exploration Agency, 1831-6 Oomagari, Kamiodagiri, Saku, Nagano 384-0306, Japan}
\altaffiltext{4}{Institute for Cosmic Ray Research, University of Tokyo, 5-1-5, Kashiwa-no-ha, Kashiwa, Chiba 277-8582, Japan}
\altaffiltext{5}{Institute of Astronomy and Astrophysics, Academia Sinica, 11F of AS/NTU Astronomy-Mathematics Building, No.1, Sec. 4, Roosevelt Rd, Taipei 10617, Taiwan, R.O.C.}
\altaffiltext{6}{Operation Division, Square Kilometre Array Observatory, Jodrell Bank Observatory, Lower Withington, Macclesfield, Cheshire SK11 9FT, UK}
\altaffiltext{7}{Japan Aerospace Exploration Agency, Institute of Space and Astronautical Science, 3-1-1, Yoshinodai, Sagamihara, Kanagawa 252-5210, Japan}
\altaffiltext{8}{Department of Space and Astronautical Science, SOKENDAI (The Graduate
University for Advanced Studies), Sagamihara
252-5210, Japan}
\altaffiltext{9}{Department of Physics, National Chung Hsing University, No. 145, Xingda Rd., South Dist., Taichung 40277, Taiwan (R.O.C.)}
\altaffiltext{10}{Department of Physics, Hiroshima University, 1-3-1, Kagamiyama, Higashi-Hiroshima, Hiroshima 739-8526, Japan}
\altaffiltext{11}{Mizusawa VLBI Observatory, National Astronomical Observatory of Japan, 1-2-2 Mizusawa-Hoshigaoka,  Oshu, Iwate 023-0861, Japan}
\altaffiltext{12}{National Institute of Information and Communications Technology, 4-2-1 Nukui-Kitamachi, Koganei, Tokyo 184-8795, Japan}
\altaffiltext{13}{Graduate School of Sciences and Technology for Innovation, Yamaguchi University, 1677-1 Yoshida, Yamaguchi, 753-8512, Japan}
\altaffiltext{14}{The Research Center for Time Studies, Yamaguchi University,  1677-1, Yoshida, Yamaguchi 753-8511, Japan}
\altaffiltext{15}{Center for Astronomy, Ibaraki University, 2-1-1, Bunkyo, Mito, Ibaraki 310-8512, Japan}
\altaffiltext{16}{Department of Physics, Graduate School of Science, Kyoto University, Kitashirakawa Oiwake-cho, Sakyo-ku, Kyoto 606-8502, Japan}
\altaffiltext{17}{Cluster for Pioneering Research, RIKEN, 2-1 Hirosawa, Wako, Saitama 351-0198, Japan}



\maketitle

\begin{abstract}
We present a detection of a bright burst from FRB 20201124A, which is one of the most active repeating FRBs, based on S-band observations with the 64-m radio telescope at the Usuda Deep Space Center/JAXA. This is the first FRB observed by using a Japanese facility. Our detection at 2 GHz in February 2022 is the highest frequency for this FRB and the fluence of $>$ $189$ Jy ms is one of the brightest bursts from this FRB source.
We place an upper limit on the spectral index $\alpha$ $=$ $-2.14$ from the detection of the S band and non-detection of the X band at the same time.
We compare an event rate of the detected burst with ones of the previous research, and suggest that the power-law of the luminosity function might be broken at lower fluence, and the fluences of bright FRBs distribute up to over 2 GHz with the power-law against frequency.
In addition, we show the energy density of the burst detected in this work was comparable to the bright population of one-off FRBs. We propose that repeating FRBs can be as bright as one-off FRBs, and only their brightest bursts could be detected so some of repeating FRBs intrinsically might have been classified as one-off FRBs.
\end{abstract}

\section{Introduction}
Fast Radio Bursts (FRBs), short-duration ($\ll$ 1 sec) bright radio transients, have still unknown progenitors and emission mechanisms \citep{Petroff2022}. Detection of FRBs at different frequencies and over wider bandwidths are helpful for studying the intrinsic emission properties and the frequency dependence of potential propagation effects due to the circumburst environment. Some proposed emission models have a limited range of radio frequency and/or spectral index (e.g., \cite{Kumar2017,Metzger2019}). Simultaneous observation covering a wide range of frequencies for FRB 20180916B discovered the frequency dependence of the activity window, which has provided potentially an important insight into the nature of these sources \citep{Pastor-Marazuela2021,Bethapudi2022}.
Some FRBs have been observed at various frequencies, such as FRB 20121102 from 400 MHz to 8 GHz, \citep{Josephy2019,Gajjar2018}, FRB 20180916B from 110 MHz to 5 GHz \citep{Chawla2020,Pleunis2021b,Bethapudi2022}.
However, in general, most of FRBs have been detected at 400 -- 800 MHz or 1.4 GHz at the rest frame.

A subset of FRBs have shown repetition, which rules out solely catastrophic progenitors. There is a possibility that one-off FRBs are detected only once due to limited observation time or sensitivity, and whether all FRBs are capable of repeating or not has been actively debated  (e.g., \cite{Ravi2019,Ai2021,Gardenier2021}). Some research (e.g., \cite{Hashimoto2020a}) propose that FRBs are classified into two populations, that is, not all FRBs repeat. It is suggested that repeating FRBs have some different trends from one-off FRBs, such as the duration time, bandwidth, redshift evolution, and especially energy; repeating FRBs tend to be fainter than one-off FRBs \citep{Hashimoto2020a,Hashimoto2020b,Pleunis2021a,Kim2022}.

The repeaters can provide an opportunity to conduct follow-up observations at higher radio frequencies.
Long-term follow-up observations over a time scale of years can constrain periodic activity.
Some repeating FRBs, such as FRB 121102 \citep{Cruces2021} and FRB 180916B \citep{CHIMEFRB2020}, have shown periodicity of the burst activity window. The repeating FRB 20201124A entered a period of high activity in April of 2021 \citep{CHIMEFRB2021,Lanman2022}, at which time several observatories recorded tens to thousands of bursts from the source. Recently, it again entered an active phase in January--March 2022 \citep{Ould-Boukattine2022a,Ould-Boukattine2022b,Atel15197}. However, no detection has been reported in the high frequency band above 2 GHz from this source.

Here we report a detection of a bright FRB from the repeater FRB 20201124A source at 2.2 -- 2.3 GHz, which is the highest-frequency detection from FRB 20201124A to date, using the 64-m radio dish of Usuda Deep Space Center (UDSC)/JAXA. This is the first FRB detected in Japan.
The remainder of this paper is structured as follows. In Section 2, we present our observational setup and analyses. In Section 3, we describe the detected burst in detail, dispersion measure (DM) determination,  digital artifact, and then estimate the spectral index. We compare the detected burst with the previous bursts from FRB 20201124A and other FRBs in Section 4, and summarize this paper in Section 5.

\section{Observations and Analyses}
\label{obs_analysis}
Following the recent reports of the 2022 reactivation of FRB 20201124A as cited in the introduction,
we conducted an observing session for this FRB source using UDSC/JAXA
for 8 hours on February 18, 2022 (MJD 59628), 07:11:00--15:14:00 UT at the S (2194 -- 2322 MHz) and X bands (8374 -- 8502 MHz).
We also observed the 3C147 as a flux calibrator before and after the FRB observation.
We used the polynomial expression for 3C147 studied by \citet{Perley2017} to speculate the flux densities in the S and X bands to be 15.2 and 4.7 Jy, respectively,
and determined values of
the system equivalent flux density (SEFD) 122 Jy (7\% uncertainty) for the S band,
and 173 Jy (40\% uncertainty) for the X band. These uncertainties resulted from atmospheric fluctuation.

We observed right-handed circular polarized waves using a multi-channel digital A/D sampler , ADS3000+ \citep{Takeuchi2006}, with a sampling rate of 64 MHz and 4 bit quantization, and divided the observed full bandwidths of 128 MHz $\times\ 2$ (S and X bands)
to 4 $\times\ 2$ channels of 32 MHz bandwidth (effectively 30 MHz width after the elimination of the band edges).
The data in each channel are coherently dedispersed with a trial value of DM = 413 pc cm$^{-3}$, which is from \citet{Xu2021},
and then incoherently summed over 4 channels to make the averaged time series for the S band and X band, separately (see \cite{Enoto2021} for more details).

\begin{figure}
 \begin{center}
 \includegraphics[width=\columnwidth]{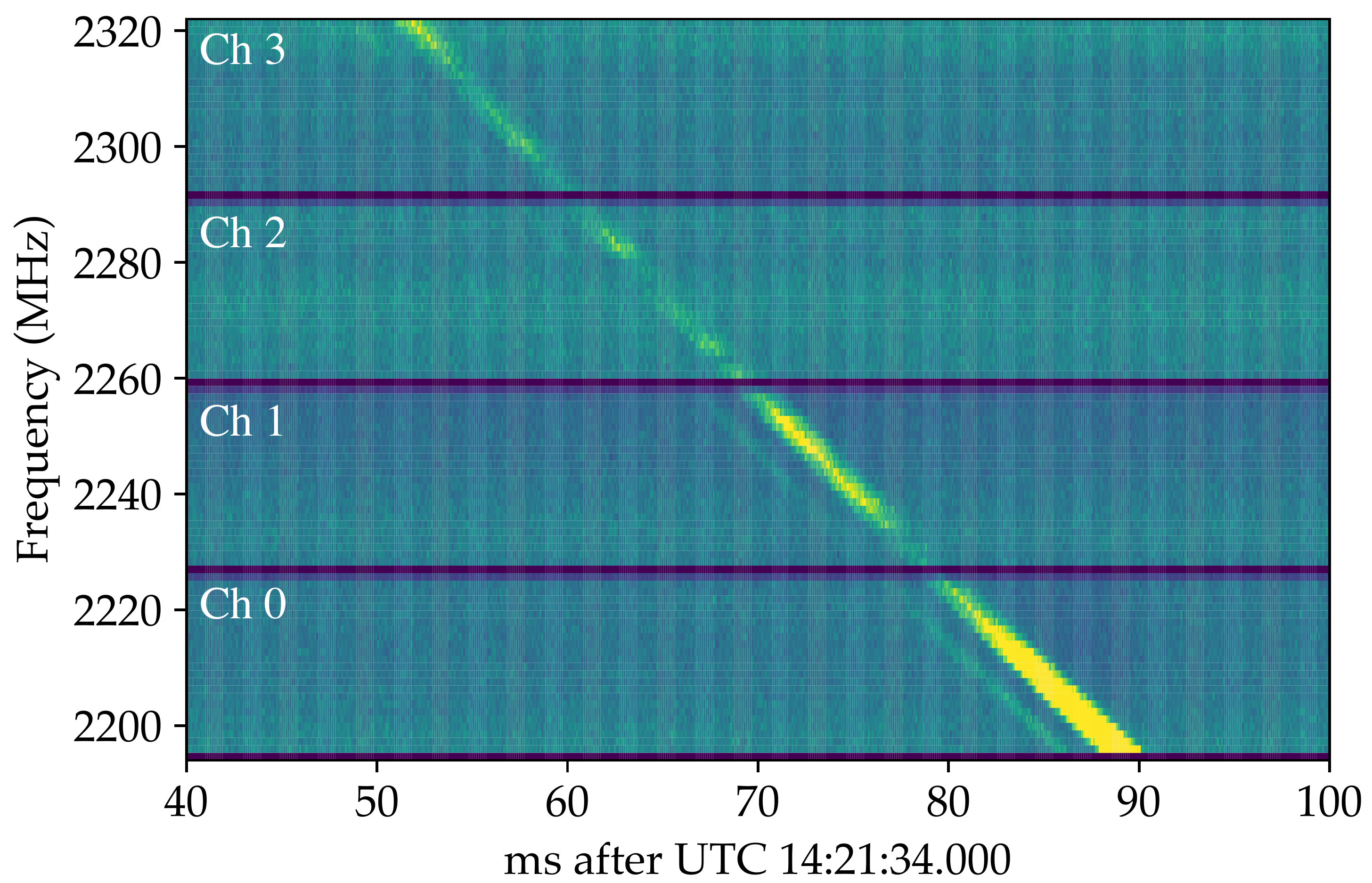} \\
 \end{center}
\caption{Dynamic spectrum of a burst from FRB 20201124A at 2 GHz band before dedispersion. Central frequencies of channel(ch) 0, ch 1, ch 2, and ch 3 are 2210, 2242, 2274, and 2306 MHz, respectively.}
\label{fig:FRB-dynamic-spectra}
\end{figure}

\begin{figure}
 \begin{center}
   \includegraphics[width=\columnwidth]{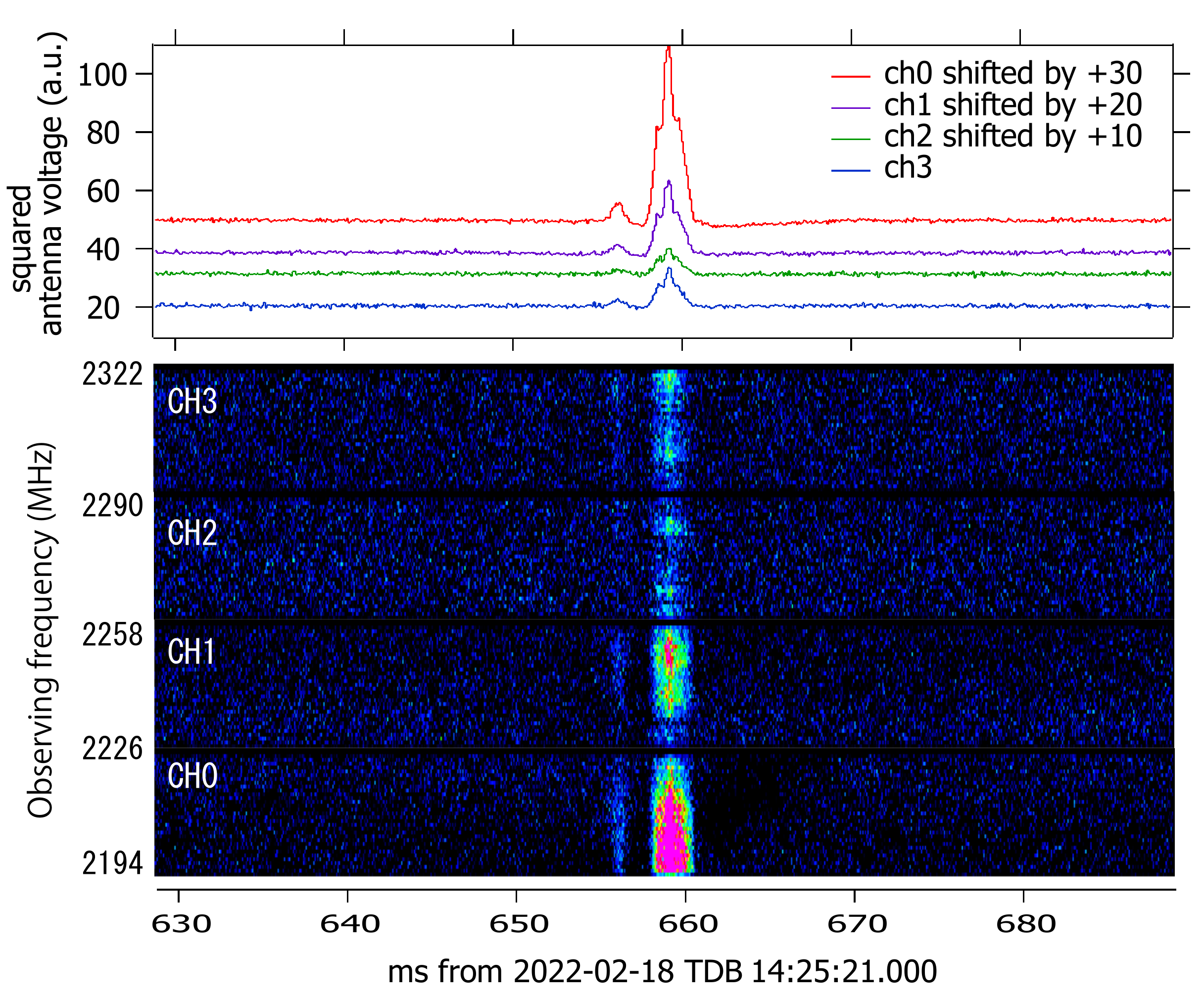}
 \end{center}
\caption{
Upper panel: Time series of squared antenna voltages at the S band with 100 $\mu$s time integration.
Lower panel: Waterfall plot of the detected FRB event after dedispersion with the best-fit DM of 411.0 pc cm$^{-3}$.
Time is converted to TDB (solar-system barycentric dynamical time) using \texttt{TEMPO2} \citep{Hobbs2006} for the infinite frequency.
The frequency resolution of the waterfall plot is 1 MHz.
}
\label{fig:waterfallplot}
\end{figure}

We searched the averaged time series of 1 ms integration for burst candidates having signal-to-noise ratios (S/N) $\ge$ 10 both in the S and X band,
namely those having peak fluxes $>$ 3.5 Jy for the S band and $>$ 5.0 Jy for the X band.

\section{Results}
After the analysis mentioned above, we found an FRB candidate having S/N $>$ 300 at the S band \citep{Atel15285} and none at the X band.
For this candidate Figure \ref{fig:FRB-dynamic-spectra} shows a dynamic spectrum before dedispersion covering a 60 ms time window starting at UTC 14:21:34.040. This dynamic spectrum shows a clear detection of a strong radio burst with a dispersive delay.
The dispersive delay apparently follows a $\nu^{-2}$ law, which is a typical characteristic of FRBs,
being caused by propagation through the interstellar/intergalactic plasma.
Intensity variation against frequency may be attributed to interstellar scintillation.

To determine the optimal DM, we averaged the dedispersed data every 100 $\mu$s and calculated the structure parameter (\cite{Gajjar2018}) defined as,
\begin{equation}
{\rm{Structure\ Parameter}} = \frac{1}{n}\sum^n_i\left|\frac{S_i-S_{i+1}}{\Delta t}\right|,
\end{equation}
where $n$ is the number of total bins which include the bursts, $S_i$ is the $i$-th flux, and $\Delta t$ is the integration time. We calculated the structure parameters by changing DM. Then we fitted the structure parameters with the Gaussian distribution and determined the structure-parameter-maximizing-DM of 411.0 $\pm$ 0.5 pc cm$^{-3}$ as the optimal DM\footnote{We searched burst candidates again by dedispersion with the optimal DM value of 411.0 pc cm$^{-3}$ and, in the end, there was only the one FRB already described in this section.}. This is consistent with values reported in the previous work within uncertainties \citep{Xu2021}.

In the upper panel of Figure \ref{fig:waterfallplot} we present the burst profiles integrated over ch 0 -- 3 separately after the dedispersion using our best estimate DM.
The lower panel of Figure \ref{fig:waterfallplot} shows the dynamic spectrum, e.g., the waterfall plot, after the dedispersion.
The bright main component has a pulse width (FWHM) of 1.5 ms which is typical for repeating FRBs \citep{Petroff2022}.
There is a faint sub-structure at 3 ms prior to the main component, with a peak intensity of $\sim$ 10\% of the peak of main component.
Similar sub-structures with a timescale of a few ms are also reported for FRB 20201124A (e.g., \cite{Marthi2022, Main2022}).

\begin{figure}[t]
 \begin{center}
  \includegraphics[width=\columnwidth]{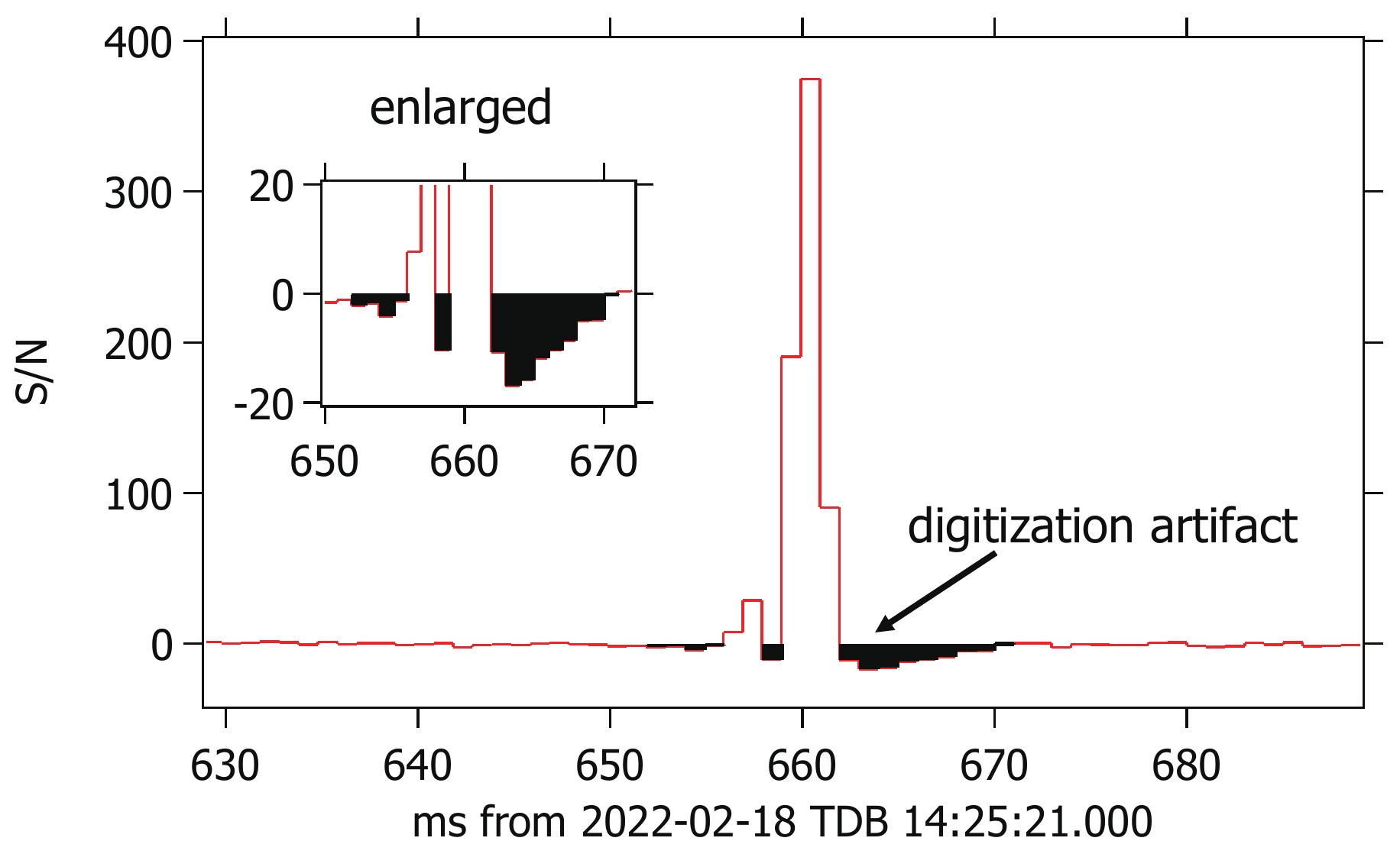}
 \end{center}
\caption{
The S/N time profile for ch 0 with the integration time of 1 ms.
The filled area shows digitization artifacts. The inset show an enlarged portion of digitization artifacts.
}\label{fig:DAeffect}
\end{figure}

We should note one caveat about the flux/fluence determination.
Since the previous observations of FRB 20201124A showed that the majority of bursts from this FRB have flux density $\lesssim$ several tens of Jy (e.g., \cite{Xu2021}),
we had optimized the digital system to this flux density level.
Figure \ref{fig:DAeffect} shows the burst profile for ch 0, and we see a dip in the noise floor in the neighboring interval of the burst peak (filled area), which represents digital artifact due to a finite bit size of the digitization system \citep{Jenet98}
and results in the flux/fluence underestimations.
For ch 1 there is also a dip although shallower than ch 0 but significant.
We see no dip for ch 2 and ch 3 and concluded that these channels are free from the digitization artifact.
Table 1 summarizes our estimation of the flux/fluence separately for ch 0 -- 3.
If we incoherently sum up these four channels, we get estimations for the peak flux $>$ 114 Jy and for fluence $>$ 189 Jy ms
for the frequency band, 2194 -- 2322 MHz. It should be noted again that only right-handed circular polarized waves were observed.
The mean value of V/I of FRBs detected in \citet{Hilmarsson2021} was -2±9 \%, where V and I are stokes parameters. From this result, the intensity of a right-handed circular polarized wave is as high as that of a left-handed circular polarized wave. Hence, FRBs that should be detected if we also carried out left-handed circular polarized wave observations have a comparable intensity of right-handed circular polarized wave, so could also be detected by this observation. To summarize, we conclude that only right-handed circular polarized wave observation would not miss FRBs, namely, would not affect the detection number. The total fluence would be higher than the value obtained from this observation.

Assuming that fluence can be expressed as $F_{\nu} \propto \nu^{\alpha}$, we calculate the spectral index of $\alpha$ for the detected burst from the simultaneous observations of the S and X bands.
We obtain the lower limit of fluence, 189 Jy ms (7 \% uncertainty),  for the S band due to the digitization artifact, and the upper limit of fluence, 7.5 Jy ms (40 \% uncertainty),  for the X band due to the non-detection by assuming that the duration in the X band is the same as that in the S band.
Considering uncertainties we get the upper limit of $\alpha$ to be $-2.14$.
This upper limit is at variance with the spectral index of the mean of bright FRBs detected with the Australian Square Kilometre Array Pathfinder (ASKAP) \citep{Macquart2019}, and magnetar emissions
(e.g., \cite{Levin2010, Eie2021}).
However it is consistent with a mean index of main pulse and interpulse for giant pulses from the Crab pulsar between 2.3 GHz and 8.4 GHz (S. Eie in preparation). This comparison could reveal clues to the emission mechanism of FRB 20201124A. Although further investigations are necessary, observations of FRB 20201124A might support the emission mechanism models like Crab giant radio pulses more.

\begin{table}[h]   \tbl{Flux/Fluence estimation for a burst of FRB 20201124A on 18 February 2022.}
{
\begin{tabular}{clll}
\hline
 ch      & ~~~~central   & ~~~flux\footnotemark[$*$] & ~~~fluence\footnotemark[$*$]         \\
         & ~~~frequency  &                           &                                      \\
\hline
  0    & 2210 MHz & $>$ 262 Jy &  $>$ 444 Jy ms\\
  1    & 2242     & $>$ 109    &  $>$ 178 \\
  2    & 2274     &  ~~~    34 &  ~~~  55 \\
  3    & 2306     &  ~~~    51 &  ~~~  79 \\
\hline
\end{tabular}
}
\begin{tabnote}
\footnotemark[$*$]7\% uncertainly
\end{tabnote}
\end{table}

\section{Discussion}
\subsection{Comparison with the previous bursts from FRB 20201124A}
During the previous active phases of FRB 20201124A, some radio telescopes, such as the Five-hundred-meter Aperture Spherical radio Telescope (FAST) and Effelsberg 100-m radio telescope, conducted follow-up observations (e.g., \cite{Xu2021,Hilmarsson2021}).
FAST detected 1863 bursts from the FRB 20201124A source from April 1 to June 11 in 2021, covering the range from 1.0 GHz to 1.5 GHz \citep{Xu2021}.
The observed fluences distributed up to $\sim$ 70 Jy ms.
\citet{Hilmarsson2021} reported 20 bursts detected with Effelsberg 100-m radio telescope at $\sim$ 1.36 GHz during observations conducted on April 9 2021.
The brightest event among the 20 bursts detected by Effelsberg 100-m radio telescope showed a fluence of $\sim$ 30 Jy ms.
Therefore, the burst detected in this work, whose flux and fluence is the lower limit determination due to the digitization artifact at $\sim$ 2 GHz,  is one of the brightest events from this FRB source (e.g., 334 Jy ms reported in \cite{Atel14556}, and $>771$ Jy ms in \cite{Ould-Boukattine2022b} at 1.3 GHz).
We also note that the detected frequency ($\sim$ 2.2 -- 2.3 GHz) is highest.

\begin{figure}[t]
 \begin{center}
  \includegraphics[width=\columnwidth]{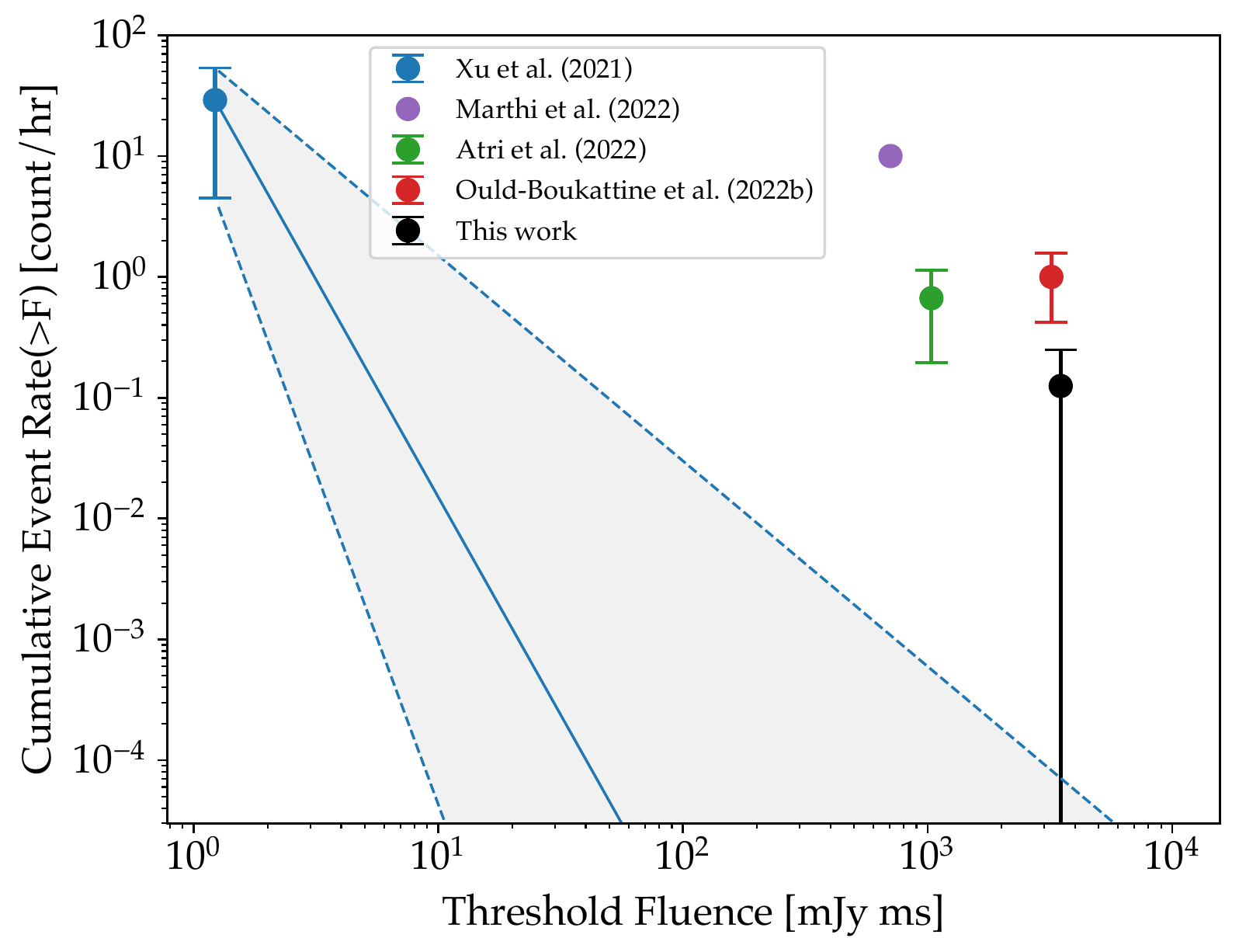}\\
 \end{center}
\caption{
Cumulative event rates of FRBs having fluence larger than threshold for each telescope. Blue, purple, green, red, black points represent cumulative event rates of \citet{Xu2021}, \citet{Marthi2022}, \citet{Atel15197}, \citet{Ould-Boukattine2022b}, this work, respectively.
\citet{Xu2021} and \citet{Marthi2022} observed last year, whereas \citet{Atel15197} and \citet{Ould-Boukattine2022b} observed this year, namely in the same active period as this work.
The blue solid line represents the event rate by extrapolating luminosity function $N (>F) \propto F^{\beta + 1}$. Gray region corresponds the range obtained from the uncertainty of $\beta$ and event rate of \citet{Xu2021}.
}\label{fig:EventRate}
\end{figure}

We contrast the event rate of FRBs speculated from the observation result of this work with the FAST result, which detected large quantities of FRBs \citep{Xu2021}.
We observed for 8 hours and detected one burst, giving an event rate of 1/8 ${\rm hr^{-1}}$. Since the observed frequency and threshold are different, we correct for these differences. Assuming all FRBs follow the form of $F_{\nu} \propto \nu^{\alpha}$ over all frequencies,
we convert the threshold of \citet{Xu2021} to the fluence at our frequency, 2258 MHz. Moreover, the single power-law of luminosity function, $N (>F) \propto F^{\beta + 1}$ where $N$ is a cumulative number per an hour, is simply assumed to extrapolate event rates for the threshold of each telescope.
In our calculation, we used $\alpha = -2.14$, which is the upper limit estimated in Section 3, and $\beta = -4.6\pm1.3\pm0.6$, which is from \citet{Lanman2022}.
Figure \ref{fig:EventRate} shows cumulative event rates of FRBs of which fluence is higher than thresholds at the same frequency.
Our event rate is notably larger than that of extrapolated from \citet{Xu2021}, $1.0 \times 10^{-11}\ {\rm hr^{-1}}$, although it is within errors of extrapolated result from \citet{Xu2021}.
We also compare the event rates from \citet{Marthi2022}, \citet{Atel15197}, and \citet{Ould-Boukattine2022b} by performing the same procedures. \citet{Marthi2022} observed last year, whereas \citet{Atel15197} and \citet{Ould-Boukattine2022b} observed this year, namely in the same active period as this work. As a consequence, the event rate extrapolated from \citet{Xu2021} is inconsistent with that of \citet{Marthi2022} even though they carried out their observations during the same active period last year. 
This suggests that the single power-law distribution of the luminosity function which we assume to estimate event rates is not correct. That is, the power-law of the luminosity function might be broken at lower fluence.
In addition, the consequence that event rates of \citet{Atel15197} and \citet{Ould-Boukattine2022b} in the same active phase we observed are roughly equivalent to our event rate implies a possibility that bright FRBs distribute up to over 2 GHz with the power-law.

\subsection{A possibility that repeating FRBs are observed as one-off FRBs }
We compare the properties of the burst of FRB 20201124A observed on 18 February 2022 (this work) to the sample collected from the FRBCAT project \citep{Petroff2016}.
More detail about the sample of FRBs is described in \url{https://www.frbcat.org/}.
FRBs are detected at different frequencies and different redshifts so, for a fair comparison, \citet{Hashimoto2020a} estimate energy densities at a unified rest-frame frequency of 1.83 GHz and rest-frame intrinsic duration.
Detailed procedures are described by \citet{Hashimoto2020a}.
After excluding FRBs without a solution to DM-derived redshift, i.e., ${\rm DM}_{\rm obs}-{\rm DM}_{\rm halo}-{\rm DM}_{\rm ISM}-{\rm DM}_{\rm host}<0$, the comparison sample includes a total of 11 repeating FRB sources with 144 repeats and 77 one-off FRBs\footnote{${\rm DM}_{\rm obs}$, ${\rm DM}_{\rm halo}$, ${\rm DM}_{\rm ISM}$ and ${\rm DM}_{\rm host}$ represent an observed DM, a DM contributed from the dark matter halo hosting the Milky Way, the interstellar medium in the Milky Way, and a FRB host galaxy.}. \citet{Hashimoto2020b} confirms that the DM-derived redshifts estimated by these procedures are consistent with spectroscopic redshifts within uncertainties (see also Figure 1).

We performed the same procedures to the FRB in this work. The lower limit of the energy density ($E_{\nu}$) of FRB 20201124A is calculated using $z_{\rm spec}=0.0979$ \citep{Fong2021}, the observed fluence of $F_{\nu}>444$ Jy ms at 2210 MHz (ch 0), and the spectral index of $\alpha<-2.14$.
The calculated lower limit is $\log (E_{\nu}/{\rm erg~Hz}^{-1})=32.24$ at 1.83 GHz.
The average fluence at 2194 -- 2322 MHz (ch 0 -- 3) is 189 Jy ms, which places the lower limits of $\log (E_{\nu}/{\rm erg~Hz}^{-1})=31.89$.
We note that the instrumental pulse broadening effects, including dispersion smearing and finite time sampling, are negligible due to the coherent dedispersion described in Section \ref{obs_analysis} and 100 $\mu$s time sampling.
Therefore, the rest-frame intrinsic duration of the burst detected in this work is $1.5/(1+z_{\rm spec})\sim1.4$ ms.

\begin{figure}[t]
 \begin{center}
  \includegraphics[width=\columnwidth]{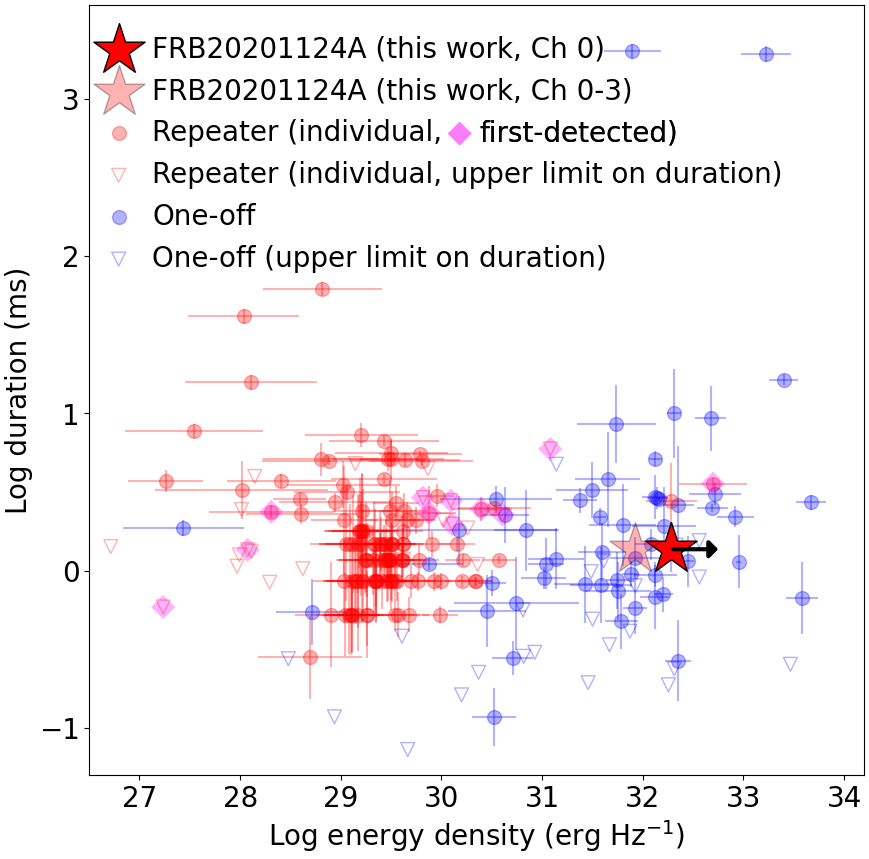}\\
 \end{center}
\caption{
Rest-frame intrinsic duration as a function of energy density at rest-frame 1.83 GHz of FRBs.
FRB 20201124A detected in this work is shown as a dark red star with an arrow.
The arrow indicates the lower limit of the energy density.
The lower limit of the energy density is calculated by the observed fluence at 2210 MHz (ch 0).
A translucent red star indicates the lower limit of the energy density derived by the fluence averaged over 2194 -- 2322 MHz (ch 0 -- 3).
The comparison samples (dots and open triangles) are collected from the FRBCAT project \citep{Petroff2016}.
Red and blue colors indicate repeating and one-off FRBs, respectively.
The repeats of each repeating FRB source are individually demonstrated.
The first-detected FRB for each repeating FRB source is highlighted in a magenta diamond.
Open triangles indicate upper limits on the rest-frame intrinsic duration whereas dots are the measured values. The other two repeating FRBs that have high energy densities comparable to that of FRB 20201124A in this work are FRB 171019, FRB 181017.J1705+68 (FRB 20181017A).
}\label{fig:comparison}
\end{figure}

Figure \ref{fig:comparison} shows the rest-frame intrinsic duration as a function of the energy density of FRB 20201124A (this work) along with those of the comparison sample. The samples of FRBs used in this work were obtained by various telescopes so include observational biases. Considering observational results, FRBs are classified into two populations, repeating and one-off FRBs. There is a possibility that this is caused by observational biases and FRBs intrinsically consist of only one population. However, some research propose that repeating and one-off FRBs have distinct physical origins. Namely, this classification a physical difference to some extent (e.g., \cite{Hashimoto2020a, ChenHY2022}).
To mitigate a possible observational bias due to targeted follow-up campaigns of repeating FRBs, the first-detected FRB for each repeating FRB source (magenta diamond in Figure \ref{fig:comparison}) can be compared with one-off FRBs (blue markers).
The first-detected repeating FRBs show lower energy densities than those of one-off events on average.
A similar result is also reported for energies of the CHIME FRB sample \citep{Kim2022}.
One possible reason is the narrower bandwidths of repeating FRBs compared to apparent one-offs (e.g., \cite{Pleunis2021a}).

The detected burst from the FRB 20201124A source shows one of the highest energy densities of repeating FRBs.
The energy density is comparable to those of the bright population of one-off FRBs.
The other two repeating FRBs show high energy densities comparable to that of FRB 20201124A; FRB 171019, FRB 181017.J1705+68 (FRB 20181017A).
The former was originally reported as one of the bright one-off FRBs detected with ASKAP at $\sim$ 1.2 GHz \citep{Shannon2018}.
Afterward, two repetitions were detected from this FRB source with the Green Bank Telescope in observations centered at 820 MHz \citep{Kumar2019}.
The two repetitions were observed as $\sim$ 590 times fainter than the ASKAP-discovered burst, indicating the importance of deep follow-up observations to identify repeating FRB sources.
Another is FRB 181017.J1705+68 (FRB 20181017A) with $\log (E_{\nu}/{\rm erg~Hz}^{-1})>32.3$.
Two repeating FRBs were detected from this source with CHIME at $\sim$500 -- 550 MHz so far \citep{CHIMEFRB2019}.
Our result highlights that repeating FRBs can be as bright as the bright population of one-off FRBs.

\citet{Ravi2019} argued that the volumetric occurrence rate of one-off FRBs likely exceeds the event rate of cataclysmic progenitor candidates, suggesting that one-off FRBs are significantly contaminated by repeating FRB sources.
Repeating FRBs are fainter than one-off events on average (e.g., \cite{Hashimoto2020a,Kim2022}).
Therefore, some repeating FRBs might have been missed due to sensitivity limitations of current radio telescopes, which would mislead the FRB classification.
The bright repeating FRB detected in this work might support the hypothesis that some of or a significant fraction of one-off events are a part of repeating bursts (e.g., \cite{Ravi2019, Kumar2019, ChenBH2022}).
However, only their brightest rare population could be detected and thus misclassified as one-off FRBs.
Monitoring the other repeating FRB sources and deep follow-up observations of one-off FRB sources are highly encouraged to prove this hypothesis.

\section{Summary}
We conducted follow-up observations of the FRB 20201124A source using the UDSC/JAXA radio telescope and detected one bright FRB. This is the first FRB detection in Japan.
The frequency of the detection, 2.2 -- 2.3 GHz, is the highest end of the multi-wavelength coverage of this repeating FRB so far reported.
In addition, the fluence at 2 GHz is comparable to the brightest events detected from this FRB source at 1.3 GHz.
By calculating the upper limit of the spectral index of the detected burst, $\alpha = -2.14$, we compare the event rate of the detected burst in this work with those of \citet{Marthi2022}, \citet{Atel15197}, \citet{Ould-Boukattine2022b}, and extrapolated from \citet{Xu2021} under the assumption of single power-law of the luminosity function and find that the event rates of the former three are close to ours and notably larger than the event rates extrapolated from \citet{Xu2021}. It is suggested that the power-law of the luminosity function might be broken at lower fluence, and bright FRBs distribute up to over 2 GHz with the power-law. In addition, from comparison of energy densities, we insist that some of the repeating FRBs intrinsically are classified as one-off FRBs due to the limited observation time or limited sensitivity ; that is, the classification based on observations may not represent the intrinsic physical mechanisms completely.

\begin{ack}
We are very grateful to the anonymous referee for many insightful comments. We thank the Usuda 64-m antenna operation support team in Space Tracking and Communication Center and ISAS, JAXA.
This work was supported in part by JSPS KAKENHI Grant Nos. 21J00416(SY), 21H01078, 22K03681 (SK), 22H01267 (YM, TE, MH, SK, YY). SY is supported by JSPS Research Fellowships for Young Scientists.
TH acknowledges the support of the National Science and Technology Council of Taiwan through grants 110-2112-M-005-013-MY3, 110-2112-M-007-034-, and 111-2123-M-001-008-.
\end{ack}


\begin{thebibliography}{}
\bibitem[Ai et al.\ (2021)]{Ai2021}
Ai, S., Gao, H., \& Zhang, B.\ 2021, \apjl, 906, L5

\bibitem[Atri et al.\ (2022)]{Atel15197}
Atri, P., Bilous, A., van Leeuwen, J., et al.\ 2022, Astron. Telegram, 15197, 1

\bibitem[Bethapudi et al.\ (2022)]{Bethapudi2022} Bethapudi, S., Spitler, L.~G., Main, R.~A, et al.\ 2022, arXiv:2207.13669

\bibitem[Chawla et al.\ (2020)]{Chawla2020} Chawla, P., Andersen, B.~C., Bhardwaj, M. \ 2020, \apj, 896, 41

\bibitem[Chen, B.~H. et al.\ (2022)]{ChenBH2022} Chen, B.~H., Hashimoto, T., Goto, T., et al.\ 2022, \mnras, 509, 1227. doi:10.1093/mnras/stab2994

\bibitem[Chen, H.-Y. et al.\ (2022)]{ChenHY2022} Chen, H.-Y., Gu, W.-M., Sun, M., et al.\ 2022, arXiv:2209.13943

\bibitem[CHIME/FRB Collaboration\ (2019)]{CHIMEFRB2019} CHIME/FRB Collaboration, Andersen, B.~C., Bandura, K., et al.\ 2019, \apjl, 885, L24. doi:10.3847/2041-8213/ab4a80

\bibitem[CHIME/FRB Collaboration\ (2020)]{CHIMEFRB2020} CHIME/FRB Collaboration, Amiri, M., Andersen, B.~C., et al.\ 2020, \nat, 582, 351

\bibitem[CHIME/FRB Collaboration\ (2021)]{CHIMEFRB2021}
CHIME/FRB Collaboration\ 2021, Astron. Telegram, 14497, 1

\bibitem[Cruces et al.\ (2021)]{Cruces2021}
Cruces, M., Spitler, L.~G., Scholz, P., et al.\ 2021 \mnras 500, 448-463

\bibitem[Eie et al.\ (2021)]{Eie2021}
Eie, S., Terasawa, T., Akahori, T., et al.\ 2021, \pasj, 73, 1563-1574

\bibitem[Enoto et al.\ (2021)]{Enoto2021}
Enoto, T., Terasawa, T., Kisaka, S., et al.\ 2021, Science, 372, 187-190,
Supplementary Material

\bibitem[Fong et al.\ (2021)]{Fong2021} Fong, W.-. fai ., Dong, Y., Leja, J., et al.\ 2021, \apjl, 919, L23. doi:10.3847/2041-8213/ac242b

\bibitem[Gajjar et al.\ (2018)]{Gajjar2018} Gajjar, V., Siemion, A. P. V., Price, D. C., et al.\ 2018, \apj, 863, 2. doi:10.3847/1538-4357/aad005

\bibitem[Gardenier et al.\ (2021)]{Gardenier2021}
Gardenier, D. W., Connor, L., van Leeuwen, J., Oostrum, L. C., \& Petroff, E.\ 2021, \aap, 647, A30

\bibitem[Hashimoto et al.\ (2020a)]{Hashimoto2020a}
Hashimoto, T., Goto, T., Wang, T.-W., et al.\ 2020, \mnras, 494, 2886-2904

\bibitem[Hashimoto et al.\ (2020b)]{Hashimoto2020b}
Hashimoto, T., Goto, T., On, A. Y. L., et al.\ 2020, \mnras, 498, 3927

\bibitem[Herrmann et al.\ (2021)]{Atel14556}
Herrmann, W.\ 2021, Astron. Telegram, 14556, 1

\bibitem[Hilmarsson et al.\ (2021)]{Hilmarsson2021} Hilmarsson, G.~H., Spitler, L.~G., Main, R.~A., et al.\ 2021, \mnras, 508, 5354. doi:10.1093/mnras/stab2936

\bibitem[Hobbs et al.\ (2006)]{Hobbs2006} Hobbs, G.~B., Edwards, R.~T., Manchester, R.~N.\ 2006, \mnras, 369, 655. doi:10.1111/j.1365-2966.2006.10302.x

\bibitem[Jenet and Anderson\ (1998)]{Jenet98}
Jenet, F. A., and Anderson, S. B.\ 1998, \pasp, 110, 1467-1478

\bibitem[Josephy et al. (2019)]{Josephy2019}
Josephy, A., Chawla, P., Fonseca, E., et al.\ 2019, \apjl, 882, L18

\bibitem[Kim et al.\ (2022)]{Kim2022} Kim, S.~J., Hashimoto, T., Chen, B.~H., et al.\ 2022, \mnras, 514, 5987. doi:10.1093/mnras/stac1689

\bibitem[Kumar et al.\ (2017)]{Kumar2017}
Kumar, P., Lu, W., \& Bhattacharya, M.\ 2017, \mnras, 468, 2726

\bibitem[Kumar et al.\ (2019)]{Kumar2019} Kumar, P., Shannon, R.~M., Os{\l}owski, S., et al.\ 2019, \apjl, 887, L30. doi:10.3847/2041-8213/ab5b08

\bibitem[Lanman et al.\ (2022)]{Lanman2022}
Lanman, A. E., Andersen, B. C., Chawla, P., et al.\ 2022, \apj, 927, 59

\bibitem[Levin et al.\ (2010)]{Levin2010}
Levin, L., Bailes, M., Bates, S., et al.\ 2010, \apjl, 721, 33-37

\bibitem[Macquart et al.\ (2019)]{Macquart2019} Macquart, J.-P., Shannon, R.~M., Bannister, K.~W., et al.\ 2019, \apjl, 872, L19. doi:10.3847/2041-8213/ab03d6

\bibitem[Main et al.\ (2022)]{Main2022}
Main, R. A., Hilmarsson, G. H., Marthi, V. R., et al.\ 2022, \mnras, 509, 3172-3180

\bibitem[Marthi et al.\ (2022)]{Marthi2022}
Marthi, V. R., Bethapudi, S. ,Main, R. A., et al.\ 2022, \mnras, 509, 2209-2219

\bibitem[Metzger et al.\ (2019)]{Metzger2019}
Metzger, B. D., Margalit, B., \& Sironi, L.\ 2019, \mnras, 485, 4091

\bibitem[Ould-Boukattine et al.\ (2022a)]{Ould-Boukattine2022a}
Ould-Boukattine, O. S., Kirsten, F., Nimmo, K., et al.\ 2022, Astron. Telegram, 15190, 1

\bibitem[Ould-Boukattine et al.\ (2022b)]{Ould-Boukattine2022b}
Ould-Boukattine, O. S., Kirsten, F., Nimmo, K., et al.\ 2022,  Astron. Telegram, 15192, 1

\bibitem[Pastor-Marazuela et al.\ (2021)]{Pastor-Marazuela2021}
Pastor-Marazuela, I., Connor, L., van Leeuwen, J., et al.\ 2021, \nat, 596, 505

\bibitem[Perley and Butler\ (2017)]{Perley2017} Perley, R. A., and Butler, B. J.\ 2017, \apjs, 230, 7

\bibitem[Petroff et al.\ (2016)]{Petroff2016} Petroff, E., Barr, E.~D., Jameson, A., et al.\ 2016, PASA, 33, e045. doi:10.1017/pasa.2016.35

\bibitem[Petroff et al.\ (2022)]{Petroff2022}
Petroff, E., Hessels, J. W. T., \& Lorimer, D. R.\ 2022, \aapr, 30, 2

\bibitem[Pleunis et al.\ (2021a)]{Pleunis2021a}
Pleunis, Z., Good, D. C., Kaspi, V. M. et al.\ 2021, \apj, 923, 1

\bibitem[Pleunis et al.\ (2021b)]{Pleunis2021b}
Pleunis, Z., Michilli, D., Bassa, C.~G. et al.\ 2021, \apj, 911, 3

\bibitem[Ravi\ (2019)]{Ravi2019} Ravi, V.\ 2019, Nature Astronomy, 3, 928. doi:10.1038/s41550-019-0831-y

\bibitem[Shannon et al.\ (2018)]{Shannon2018} Shannon, R.~M., Macquart, J.-P., Bannister, K.~W., et al.\ 2018, \nat, 562, 386. doi:10.1038/s41586-018-0588-y

\bibitem[Takefuji et al.\ (2022)]{Atel15285}
Takefuji, K., Murata, Y., Ikebe, S., et al.\ 2022, Astron. Telegram, 15285, 1

\bibitem[Takeuchi et al.\ (2006)]{Takeuchi2006}
Takeuchi, H., Kimura, M., Nakajima, J., et al.\ 2006, \pasp, 850, 1739-1748

\bibitem[Xu et al.\ (2021)]{Xu2021} Xu, H., Niu, J.~R., Chen, P., et al.\ 2021, arXiv:2111.11764

\end{thebibliography}
\end{document}